\documentclass[%
 reprint,
 amsmath,amssymb,
 twocolumn,
longbibliography,
floatfix,
superscriptaddress,
]{revtex4-2}

\usepackage[english]{babel}
\usepackage{graphicx}
\usepackage{dcolumn}
\usepackage[utf8]{inputenc} 
\usepackage{bm}
\usepackage{hyperref}
\usepackage{mathrsfs}
\usepackage{longtable}
\usepackage{multirow,booktabs,dcolumn,tabularx}
\usepackage{float}

\usepackage{braket}
\usepackage{microtype} 
\usepackage[dvipsnames]{xcolor}

\usepackage{amsmath,amsfonts,amsthm} 


\usepackage{dsfont}
\usepackage{comment}
\usepackage{physics}
\usepackage{booktabs}
\usepackage{pifont}
\usepackage{makecell}

\definecolor{limegreen}{rgb}{0.2, 0.8, 0.2}
\definecolor{orange}{rgb}{1.0, 0.5, 0.0}

\definecolor{emerald}{rgb}{0.31, 0.78, 0.47}
\definecolor{blue(ncs)}{rgb}{0.0, 0.53, 0.74}

\begin{document}

\preprint{APS/123-QED}

\author{Jose Antonio Moreno}
\affiliation{Laboratorio de Bajas Temperaturas y Altos Campos Magn\'eticos, Departamento de F\'isica de la Materia Condensada, Instituto Nicol\'as Cabrera and Condensed Matter Physics Center (IFIMAC), Unidad Asociada UAM-CSIC, Universidad Aut\'onoma de Madrid, E-28049 Madrid, Spain}

\author{Pablo Garc\'ia Talavera}
\affiliation{Laboratorio de Bajas Temperaturas y Altos Campos Magn\'eticos, Departamento de F\'isica de la Materia Condensada, Instituto Nicol\'as Cabrera and Condensed Matter Physics Center (IFIMAC), Unidad Asociada UAM-CSIC, Universidad Aut\'onoma de Madrid, E-28049 Madrid, Spain}

\author{Edwin Herrera}
\affiliation{Laboratorio de Bajas Temperaturas y Altos Campos Magn\'eticos, Departamento de F\'isica de la Materia Condensada, Instituto Nicol\'as Cabrera and Condensed Matter Physics Center (IFIMAC), Unidad Asociada UAM-CSIC, Universidad Aut\'onoma de Madrid, E-28049 Madrid, Spain}

\author{Sara L\'opez Valle}
\affiliation{Laboratorio de Bajas Temperaturas y Altos Campos Magn\'eticos, Departamento de F\'isica de la Materia Condensada, Instituto Nicol\'as Cabrera and Condensed Matter Physics Center (IFIMAC), Unidad Asociada UAM-CSIC, Universidad Aut\'onoma de Madrid, E-28049 Madrid, Spain}

\author{Zhuoqi Li}
\affiliation{Department of Physics and Astronomy, Iowa State University, Ames, Iowa 50011, USA}
\affiliation{Ames National Laboratory, Iowa State University, Ames, Iowa 50011, USA}

\author{Lin-Lin Wang}
\affiliation{Department of Physics and Astronomy, Iowa State University, Ames, Iowa 50011, USA}
\affiliation{Ames National Laboratory, Iowa State University, Ames, Iowa 50011, USA}

\author{Sergey Bud'ko}
\affiliation{Department of Physics and Astronomy, Iowa State University, Ames, Iowa 50011, USA}
\affiliation{Ames National Laboratory, Iowa State University, Ames, Iowa 50011, USA}

\author{Alexander I. Buzdin}
\affiliation{University Bordeaux, LOMA UMR-CNRS 5798, F-33405 Talence Cedex, France}

\author{Isabel Guillam\'on}
\affiliation{Laboratorio de Bajas Temperaturas y Altos Campos Magn\'eticos, Departamento de F\'isica de la Materia Condensada, Instituto Nicol\'as Cabrera and Condensed Matter Physics Center (IFIMAC), Unidad Asociada UAM-CSIC, Universidad Aut\'onoma de Madrid, E-28049 Madrid, Spain}

\author{Paul C. Canfield}
\affiliation{Department of Physics and Astronomy, Iowa State University, Ames, Iowa 50011, USA}
\affiliation{Ames National Laboratory, Iowa State University, Ames, Iowa 50011, USA}

\author{Hermann Suderow}
\affiliation{Laboratorio de Bajas Temperaturas y Altos Campos Magn\'eticos, Departamento de F\'isica de la Materia Condensada, Instituto Nicol\'as Cabrera and Condensed Matter Physics Center (IFIMAC), Unidad Asociada UAM-CSIC, Universidad Aut\'onoma de Madrid, E-28049 Madrid, Spain}

\title{Robust two-dimensional surface superconductivity and vortex lattice in the Weyl semimetal $\gamma$-PtBi$_2$}

	\begin{abstract}
The layered compound $\gamma$-PtBi$_2$ is a topological semimetal with Fermi arcs at the surface joining bulk Weyl points. Recent work has found signatures of surface superconductivity consisting of gap openings compatible with a critical temperature orders of magnitude larger than the bulk value. However, no superconducting vortices have been identified, raising questions about the robustness of the phase coherence. Here we use very low temperature Scanning Tunneling Microscopy (STM) and find robust superconductivity with T$_C=$2.9 K and H$_{C2}\approx$1.8 T linked to the Fermi arcs. We observe quantized superconducting vortices and the Josephson effect, demonstrating two-dimensional macroscopic quantum phase coherence.
	\end{abstract}

\maketitle

Many semimetals present Weyl or Dirac nodes in their electronic band structures\,\cite{annurevconmatphys031016025458,RevModPhys90.015001}. The bulk-boundary correspondence often dictates the presence of topologically protected bound states at the surface, which are of particular interest in superconductors because of possible non-Abelian modes\,\cite{Volovik1999}. Nevertheless, few of these semimetals are superconductors in the bulk. This has lead to proposals to study heterostructures consisting of a thin layer of a topological semimetal on top of a bulk superconductor\,\cite{PhysRevLett.100.096407,PhysRevLett.122.146803,RevModPhys90.015001,annurevconmatphys030212184337,Alicea2012,Sun2017}. Superconductivity is then induced in the surface state of the semimetal by proximity. Contrasting these proposals, it could also happen that superconductivity is intrinsically realized on the very same topologically protected surface state of a bulk material with Dirac or Weyl nodes\,\cite{Nomani2023,Bai2025,maeland2025}. Recent experiments and calculations suggest that this situation could be realized in the layered Weyl semimetal $\gamma-$PtBi$_2$\,\cite{PRB91205128,Kuibarov2024,Zabala2024,bashlakov2022,Schimmel2024,guo2025,Hoffmann2024}. The realization of such surface two-dimensional superconductivity with topologically protected states could have important applications in quantum devices\,\cite{Read2000,Nayak2008,Sato2017} and complement efforts to understand two-dimensional superconductivity in rhombohedral, twisted bilayer and trilayer graphene\,\cite{Cao2018,Han2025,mahapatra2025quantumcriticalitytunablegriffiths}, few-layer single crystals\,\cite{Kezilebieke2020,Deng2024} and ultra-thin films\,\cite{PhysRevB.103.214512,Lechermann2024,fan2025,Ming2023}.
 
$\gamma-$PtBi$_2$ crystallizes in a layered trigonal structure without inversion symmetry\,\cite{shipunov2020}. The band structure contains wide bands crossing the Fermi level at the upper and lower borders of the Brillouin zone. However, the center of the Brillouin zone is gapped. This provides a semimetallic character presenting a huge and non-saturating magnetoresistance\,\cite{Gao2018}. The electronic band structure contains Weyl points at approximately 50~meV above the Fermi level\,\cite{annurevconmatphys031016025458,RevModPhys90.015001,Gao2018}. The calculated electronic band structure agrees well with the angle resolved photoemission (ARPES), quasiparticle scattering (QPI) and quantum oscillations experiments\,\cite{Hoffmann2024,Kuibarov2024,oleary2025,PhysRevResearch.2.022042}. Furthermore, ARPES finds Fermi arcs in surface bands located close to the projection of the Weyl points to the surface plane\,\cite{Kuibarov2024,oleary2025}. Specific heat of bulk $\gamma-$PtBi$_2$ does not present traces of superconductivity down to 0.5~K. There are large resistive transitions with $T_c$ well below 1~K in thin flakes of $\gamma-$PtBi$_2$ obtained by exfoliation\,\cite{Schimmel2024,Veyrat2023,Zabala2024,shipunov2020,wang2021}. Other indications for surface superconductivity present $T_c$'s well above 10~K, associated with superconducting gap values of several tens of meV and are based on ARPES and Scanning Tunneling Microscopy (STM) measurements\,\cite{Kuibarov2024,bashlakov2022,Schimmel2024,guo2025,Hoffmann2024}. The values of the superconducting gap obtained from ARPES and STM vary widely among different experiments, with recent ARPES measurements showing no signature of a superconducting gap opening on the Fermi arcs at temperatures of 3\,K and within an energy range down to below 1~meV\,\cite{Schimmel2024,oleary2025}. What is more, the gaps observed around the Fermi level in STM often persist up to very strong magnetic fields (up to 12~T) and there are no reported observations of a vortex lattice nor of the Josephson effect\,\cite{Schimmel2024,besproswanny2025temperaturedependencesurfacesuperconductivity}. To better understand the eventual connection of superconductivity and Fermi arcs, it is important to demonstrate the presence of vortices and thus flux quantization in the surface superconductivity of $\gamma-$PtBi$_2$. 

Here, we use dilution refrigerator STM under magnetic fields to study $\gamma-$PtBi$_2$. We find homogeneous superconductivity down to atomic scale with a critical temperature of $T_c=$2.9~K and the gap size and temperature dependence following BCS theory. We observe superconducting vortices and study the behavior of the vortex lattice as a function of magnetic field and temperature. QPI shows features of the Fermi arcs at the superconducting quasiparticle peaks.

\begin{figure*}[ht!]
	\centering
		\includegraphics[width=0.8\linewidth]{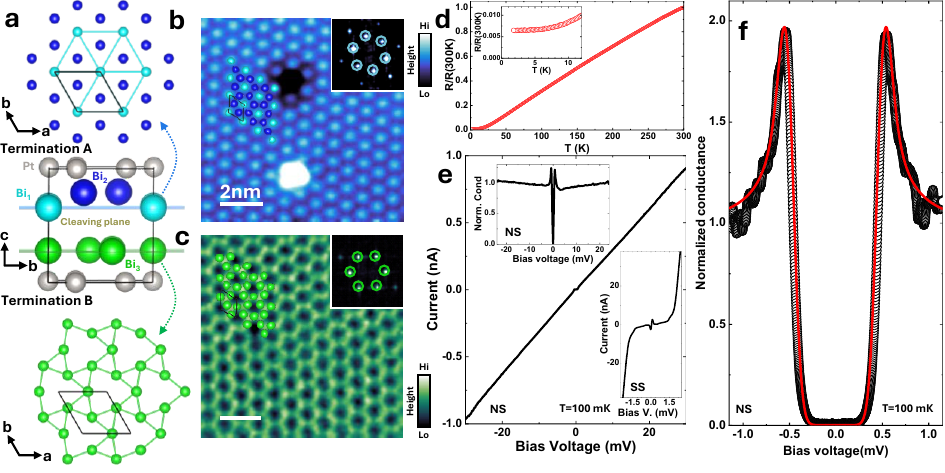} 
	\caption{
    (a) Unit cell of $\gamma$-PtBi$_2$ (space group No. 157), with Bi atoms in blue and green and Pt atoms in grey. The system is composed of layers where Pt is surrounded by Bi, so that the surface after cleaving is always composed of Bi. (b,c) Atomically-resolved STM topography taken on the two possible terminations of $\gamma$-PtBi$_2$. White scale bar corresponds to $2$ nm. Colormap is shown to the right. Inset shows the FFT with the Bragg peaks marked with circles. The atomic arrangement of Bi atoms is schematically shown by colored circles in the main panels. The surface unit cell is shown as a black rhombus. Notice that the hexagonal pattern observed in the STM image for termination A (b) is formed by the Bi$_1$ atoms. The Bi$_2$ atoms are located farther below the surface, as schematically shown in (a). The hexagonal pattern observed in the STM image for termination B (c) is formed by groups of three Bi$_3$ atoms that lie at the surface plane. These two surfaces are similar to the ones found previously in Ref.\,\cite{Schimmel2024}, although, as shown below, the tunneling conductance found here on these surfaces is very different. (d) Resistivity measurement as a function of temperature on $\gamma$-PtBi$_2$ (red circles). (e) Current vs voltage curve obtained with a normal (PtIr) tip. Top inset shows conductance vs voltage curve in the same voltage range ($T=0.1$\,K, $V=30$\,mV, and $I=1$~nA). Bottom inset shows a current vs voltage curve obtained with a superconducting (Pb) tip. We observe a sharp zero bias signal showing the Josephson effect ($T=0.1$\,K, $V=3.5\,$mV, $I=90$~nA). (f) Tunneling conductance vs bias voltage obtained with the same parameters as in (e) and $V=1.2$\,mV. Red line is a fit to a modified BCS density of states described in the text and in Fig.\,\ref{fig:gapvst}.}
		\label{fig:intro}
\end{figure*}

\begin{figure*}[ht!]
	\centering
		\includegraphics[width=0.8\linewidth]{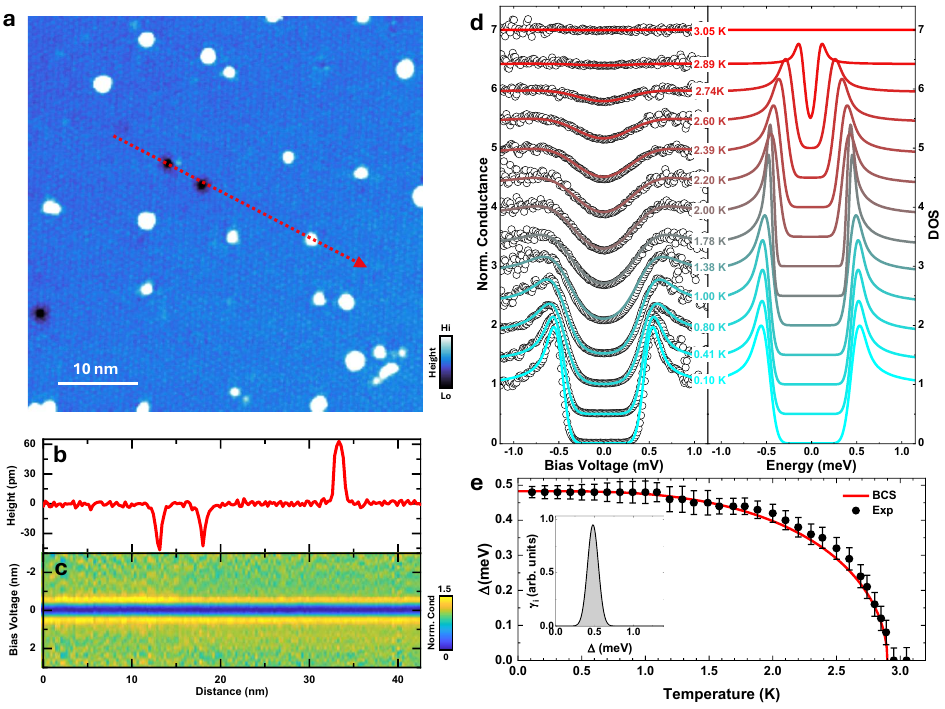} 
	\caption{(a) Atomically-resolved STM topography presenting a few defects on termination A. White scale bar corresponds to $10$~nm. Colormap is shown to the right. (b) Profile along the red arrow on the topography shown in (a). (c) Conductance as a function of bias voltage taken simultaneously as the topography, following the same profile as (b). Colormap is shown to the right. (d) On left panel we show the tunneling conductance vs bias voltage as a function of temperature (dots) and calculated conductance curves obtained after convoluting the BCS DOS (shown on the right panel) with the derivative of the Fermi function at each temperature, as described in the text. Curves are shifted vertically for clarity. (e) Temperature dependence of the superconducting gap, obtained as described in the text. The solid line is the BCS temperature dependence of the superconducting gap. Inset shows the Gaussian distribution $\gamma_i(\Delta_i)$ centered around $\Delta_0=0.48$~meV with a width of 0.07~meV used for the T=100~mK fit.}
		\label{fig:gapvst}
\end{figure*}

 The cleaving plane of $\gamma$-PtBi$_2$ is located in between two Bi layers, as shown in Fig.\,\ref{fig:intro}(a). The two possible surfaces present a trigonal arrangement of Bi atoms, one in which Bi atoms located at two nonequivalent sites (Bi$_1$ and Bi$_2$, termination A) form intertwined corrugated hexagons, and another where the Bi are located on a single plane (Bi$_3$, termination B). The resulting STM topographic images show hexagonal patterns (Fig.\,\ref{fig:intro}(b,c)), characteristic of each termination, as found previously in Ref.\,\cite{Schimmel2024}. Both terminations provide the same superconducting and vortex lattice properties described below.

In Fig.\,\ref{fig:intro}(e) we show a characteristic current vs voltage curve for $\gamma$-PtBi$_2$ at 0.1~K obtained on atomically flat areas. We note that the current is linear as a function of the voltage, suggesting excellent metallic behavior. We also note the clear and small superconducting gap that opens and a vanishing current inside the gap. Furthermore, when we use a superconducting Pb tip, we find a sharp signal at zero voltage (bottom inset in Fig.\,\ref{fig:intro}(e)) showing the Josephson effect. Further details are provided in the Supplemental Material \,\cite{SM_1}. To better visualize the superconducting gap, we show the tunneling conductance vs bias voltage in Fig.\,\ref{fig:intro}(f). We find large quasiparticle peaks and vanishing density of states inside the gap. We fit the tunneling conductance using a modified BCS density of states $DOS(E)$ with a gaussian distribution of gap values $\gamma_i(\Delta_i)$ (see End Matter). We find a Gaussian distribution centered around $\Delta_0=0.48$~meV with a width of 0.07~meV.

The spatial dependence of the superconducting tunneling conductance at atomic scale is shown in Fig.\,\ref{fig:gapvst}(a-c). The topography shown in Fig.\,\ref{fig:gapvst}(a) provides an atomically flat surface, containing a few defects, which appear as atomic size adatoms or interstitials (white spots) and as atomic vacancies (black spots). In Fig.\,\ref{fig:gapvst}(b) we represent a line scan of the height measured by STM. We see that the atomic corrugation is small, of a few pm. At the atomic size defects, there are peaks or troughs in the surface of a few tens of pm. Along the whole path, the superconducting gap slightly fluctuates (Fig.\,\ref{fig:gapvst}(c)). The defects do not strongly influence the tunneling conductance.

The temperature dependence of the superconducting tunneling conductance is shown in Fig.\,\ref{fig:gapvst}(d,e). We can fit the whole temperature dependence using the $DOS(E)$ discussed previously, convoluted by the derivative of the Fermi function at the temperature of the measurement, without any adjustment parameters up to about 1~K. Above that temperature, the value of the peak is reduced in $\gamma_i(\Delta_i)$, which gives the $DOS(E)$ provided in the right panel of Fig.\,\ref{fig:gapvst}(d). We plot the peak position in $\gamma_i(\Delta_i)$ as a function of temperature in Fig.\,\ref{fig:gapvst}(e) and find an excellent agreement with BCS theory (red line), obtained by solving the self-consistent gap equation with T$_c=2.9$~K and $\Delta_{BCS}=1.76\,k_B\,T_c=0.44$~meV. The latter coincides with $\Delta_0$ better than within 10\%.

Whereas we observe superconductivity for temperatures below 3~K, compatible with the recent ARPES experiments of Ref.\,\cite{oleary2025} and the STM work of Ref.\,\cite{Zhang25}, the temperature range for earlier claims of superconductivity in $\gamma$-PtBi$_2$ is reported to be even far above the T$_c$ we observe here\,\cite{Kuibarov2024,Zabala2024,bashlakov2022,Schimmel2024,guo2025,10.1063/5.0272618}. The repeated observation of the same gap values in our experiment shows that the superconducting properties observed here are reproducible. As we have now shown a superconducting state which exists on the surface and is robust, it is also interesting to ask about the relationship with the topological properties of the electronic band structure.

\begin{figure*}[ht!]
	\centering
		\includegraphics[width=0.8\linewidth]{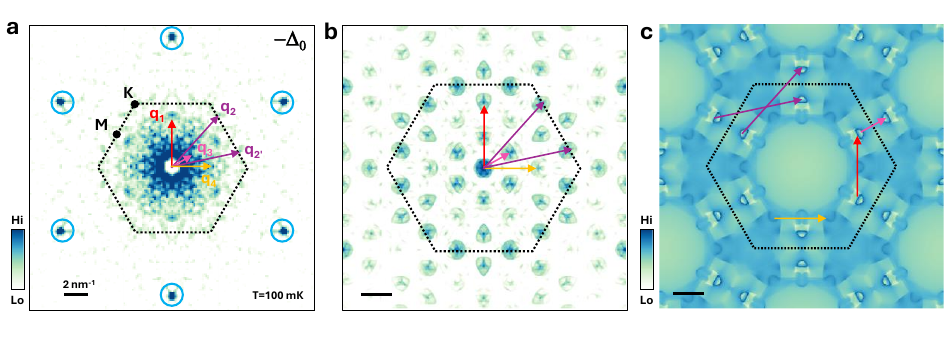} 
	\caption{(a) Quasiparticle interference pattern in $\gamma-$PtBi$_2$ measured on the region shown in Fig. \ref{fig:gapvst} (a) at 0.1~K. The scattering intensity is shown on a color scale, following the bar on the left. The dashed line is the first Brillouin zone. K and M denote high symmetry points. We mark the position of Bragg peaks with blue circles. We mark relevant scattering wavevectors $q_1$, $q_{2}$,  $q_{2'}$,  $q_{3}$ and  $q_{4}$ by colored arrows. The pattern is obtained from a tunneling conductance map at the bias voltage of the BCS quasiparticle peak, V=-$\Delta_0/e=-0.48$\,mV. (b) Joint density of states obtained from density functional theory (DFT) calculations of the the Fermi surface (shown in (c)). The colored arrows are the scattering wave vectors found in (a). (c) Fermi surface of $\gamma-$PtBi$_2$. The Brillouin zone is shown by a dashed black line in (b,c).}
		\label{fig:QPIv2}
\end{figure*}

To investigate the relationship between surface two-dimensional superconductivity and the Fermi arcs found in ARPES\,\cite{Kuibarov2024,oleary2025}, we discuss now QPI experiments made in small areas with atomic resolution. We show the Fourier transform of the tunneling conductance map at $V=-\Delta_0/e$ in Fig.\,\ref{fig:QPIv2}(a). A similar result is found for $V=+\Delta_0/e$. We find several scattering wave vectors agreeing with previous QPI and ARPES experiments at energies above the superconducting gap edge\,\cite{Hoffmann2024,oleary2025}. By comparing with Density Functional Calculations (DFT, Fig.\,\ref{fig:QPIv2}(b,c)) we find, at the superconducting gap edge, scattering wave vectors ($q_1$, $q_2$, $q_2'$ and $q_3$) which coincide with wave vectors joining Fermi arcs that connect the projections of Weyl points to the surface.

\begin{figure*}[ht!]
	\centering
		\includegraphics[width=0.85\linewidth]{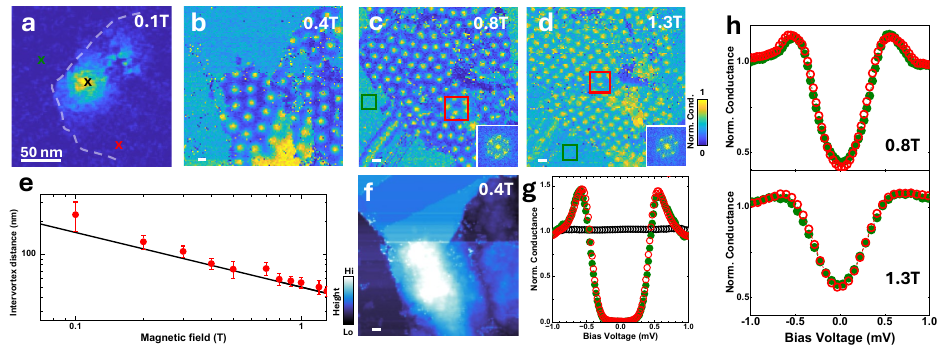} 
	\caption{(a) Zero-bias tunneling conductance map of a single vortex at 0.1\,T. White dashed line shows the boundary between an atomically flat surface on the left and a nanometer-sized flakes (see Supplemental Material\,\cite{SM_1}). (b,c,d) Tunneling conductance maps at zero bias showing the vortex lattice as a function of the magnetic field at 0.4\,T (b), 0.8\,T (c), and 1.3\,T (d). The maps shown in (c,d) were taken on the same field of view. White scale bars correspond to $50$\,nm in the main panels. Insets in (c,d) on the bottom right corner show the Fourier transform. (e) Intervortex distance as a function of the magnetic field is shown as red disks. The error bars are obtained by Delaunay triangulating vortex positions and measuring the width of the distance distribution. Solid line corresponds to the Abrikosov prediction for an hexagonal vortex lattice $d_{vortex}\sim1.075 \left(\frac{\Phi_0}{B}\right)^{1/2}$. (f) Topography taken simultaneously to the conductance map shown in (b). (g) Normalized tunneling conductance curves obtained at the core of the vortex (black dots, black cross in (a)) and away from the vortex (red and green cross, taken far from the vortex core in an atomically flat region, green and on a flake, red). (h) Top (bottom): Average of conductance curves over the regions marked with rectangles in (c,d) are shown by circles: in red the regions presenting a vortex lattice and in green the regions presenting no apparent vortex lattice in STM measurements at 0.8~T (1.3~T). Data taken at 0.1~K.}
		\label{fig:vortices}
\end{figure*}

We present the superconducting vortex lattice, obtained by mapping the zero bias conductance as a function of the position over large size areas, for different values of the magnetic field and at 0.1~K in Fig.\,\ref{fig:vortices}. In Fig.\,\ref{fig:vortices}(a) we show a zero bias conductance map with a single vortex (see Supplemental Material\,\cite{SM_1}). In Fig.\,\ref{fig:vortices}(b,c,d) we show zero bias tunneling conductance maps obtained at 0.4~T, 0.8~T and 1.3~T respectively. The intervortex distance as a function of the magnetic field (Fig.\,\ref{fig:vortices}(e)) follows Abrikosov behavior for a hexagonal vortex lattice\,\cite{RevModPhys.66.1125,Brandt1995}. The Fourier transforms of vortex lattice images mostly present a hexagonal patterns, although the vortex lattice is disordered (insets of Fig.\,\ref{fig:vortices}(c,d)). Interestingly, we find extended areas where we do not observe vortices under magnetic fields, see left part of Fig.\,\ref{fig:vortices}(a) and portions of Fig.\,\ref{fig:vortices}(b,c,d). Previous work focused on such areas and did not find indications for a vortex lattice\cite{Schimmel2024,Hoffmann2024,Zhang25}. In Fig.\,\ref{fig:vortices}(f) we show the topography of the field of view shown in Fig.\,\ref{fig:vortices}(b), see Supplemental Material for topographies in other fields of view. The areas where we do not observe vortices are atomically flat, and the areas where we observe vortices present a surface with a corrugation of the order of nanometers. The superconducting tunneling conductance is equivalent in both areas. We show the tunneling conductance far from a single isolated vortex in both areas in Fig.\,\ref{fig:vortices}(g) (red and green circles). Furthermore, the superconducting tunneling conductance averaged over a single unit cell of the vortex lattice in the areas where we observe vortices is equivalent to the tunneling conductance found in the areas where we do not observe vortices (Fig.\,\ref{fig:vortices}(h)). We provide further line scans in the Supplemental Material. All this, together with scanning electron microscopy structural and elemental characterization provided in Supplemental Material \,\cite{SM_1}, show that the areas with an irregular topography are nanometer-sized flakes of the same sample produced during the cleave.

We do not have a clear answer to the apparent absence of vortices in these atomically flat surfaces and the results found at high magnetic fields in Refs.\cite{Schimmel2024,Hoffmann2024,Zhang25}. The lack of observable vortices does not necessarily preclude their presence, however. We hypothesize effects related to enhanced vortex mobility through thermal excitation\cite{Guillamon2009,vinokur90} or tip-sample interaction\,\cite{Blatter96} and discuss further experiments in the Supplementary Material. In the areas where we do observe vortices we find (i) enhanced vortex mobility, (ii) thermally activated vortex motion, leading to vortex lattice melting when increasing temperature well below T$_c$, and (iii) little or no dependence of the vortex density from the magnetic field history, suggesting a very large magnetic field penetration. These findings (described in detail in the Supplementary Material) are hallmarks of two-dimensional vortex lattices\,\cite{Guillamon2009,PhysRevB.108.L180503,Duhan2025,Suderow_2014,Ge2016,10.1063/1.1754056,Brandt1995,RevModPhys.66.1125} and suggest that we could be observing two-dimensional vortices at the surface of $\gamma$-PtBi$_2$.

 We have thus established that robust two-dimensional superconductivity with $T_c=2.9$~K exists on the surface of $\gamma$-PtBi$_2$. We observe the Josephson effect when using superconducting tips (see Supplemental Material\,\cite{SM_1}), implying a well-defined phase difference across the junction. Furthermore, the Cooper pair wave function is uniquely valued, as shown by the observation of flux quantization and the vortex lattice.

Our QPI measurements show that the superconducting gap opens at the Fermi arcs, in agreement with theoretical predictions\,\cite{PhysRevB.110.054504}. For superconducting Fermi arcs that are projections of bulk Weyl points, Majorana modes are expected inside vortices. These should be visible by a peak in the tunneling conductance inside the vortex core\,\cite{Volovik1999,PhysRevLett.100.096407,PhysRevLett.122.146803,RevModPhys90.015001,annurevconmatphys030212184337,Alicea2012,Sun2017}. But even in absence of Majorana modes, superconductors in the clean limit present a large density of states inside vortex cores\,\cite{CAROLI1964307,PhysRevLett.64.2711,PhysRevLett.101.166407,Suderow_2014,Park2021,PhysRevLett.80.2921,PhysRevLett.67.1650}. We do find a flat tunneling conductance inside vortex cores which suggests that the electronic mean free path on nanometer-sized flakes is comparable to or smaller than the coherence length ($\xi_{a,b}\sim$13 nm), smearing localized vortex core states despite surface superconductivity being present. 
In contrast, vortices on atomically flat surfaces are most likely in the clean limit and may host bound states, potentially including Majorana modes \,\cite{maeland2025}.

In summary, we have found robust surface superconductivity in $\gamma$-PtBi$_2$ connected to the Fermi arcs joining the projection of bulk Weyl points. An interesting prospect is to better understand the pairing mechanism leading to two-dimensional surface superconductivity. Changes in the phonon structure close to the surface could play a role\,\cite{Nomani2023,Bai2025}. Interestingly, monolayer $\gamma-$PtBi$_2$ has been proposed to be a ferroelectric metal due to phonon softening \,\cite{PhysRevLett.133.186801}. Furthermore, it has been recently proposed that metals with Weyl points can show topological surface superconductivity\,\cite{maeland2025}. Superconductivity could be favored by the enhanced electronic density of surface states, in a distinct but somewhat comparable situation to the flat band surface state in rhombohedral graphene\,\cite{Han2025,jiang2025idealquantumgeometrysurface,Heikkila2011,dsouza2026}. $\gamma$-PtBi$_2$ provides an exceptional case of two-dimensional superconductivity arising at the surface state of a topological semimetal.

\par 
\par

\section*{Acknowledgments}
\noindent This work was conceived during the visit of JA Moreno to Ames Lab. We acknowledge fruitful discussions with Miguel \'Agueda, Isidoro Poveda, and Adam Kaminski. We acknowledge support by the Spanish Research State Agency (PID2020-114071RB-I00, PID2023-150148OB-I00, TED2021-130546B\-I00, PDC2021-121086-I00 and CEX2023-001316-M), the European Research Council PNICTEYES through grant agreement 679080 and VectorFieldImaging Grant Agreement 101069239, the EU through grant agreement No. 871106 and by the Comunidad de Madrid through projects TEC-2024/TEC-380 “Mag4TIC”, and SI4/PJI/2024-00199 “QUASURF”. We have benefited from collaborations through EU program Cost CA21144 (superqumap), and from SEGAINVEX at UAM in the design and construction of STM and cryogenic equipment. Work done at Ames National Laboratory (PCC, SLB, LLW, ZL) was supported by the U.S. Department of Energy, Office of Basic Energy Science, Division of Materials Sciences and Engineering. Ames National Laboratory is operated for the U.S. Department of Energy by Iowa State University under Contract No. DE-AC02-07CH11358. First-principles calculations have used resources of the National Energy Research Scientific Computing Center (NERSC), a DOE Office of Science User Facility.

\newpage
\section*{End Matter}
\subsection{Experimental details}
High-quality single crystals of $\gamma-$PtBi$_2$ were grown using a Bi-rich flux\,\cite{oleary2025}, as described in the Supplemental Material\,\cite{SM_1}. When probed macroscopically either by resistivity or by magnetization, there are no hints of bulk superconductivity down to $1.8$~K (see Fig.\,\ref{fig:intro}(d) and Supplemental Material\,\cite{SM_1}). The residual resistance ratio is of RRR$\sim245$, showing that the crystals have minimal disorder scattering. We used a dilution-refrigerator-STM with an energy resolution below 8 $\mu$eV cooling to $\sim$0.1\,K, described in Refs.\,\cite{Suderow2011,Fernandez2021}, and the data treatment software described in Refs.\,\cite{Fran2021,Horcas07,githublbtuam}. Further details on the STM data acquisition are described in the Supplemental Material\,\cite{SM_1}.
\subsection{Temperature convolution}
We fit the tunneling conductance using a modified BCS density of states with $DOS(E)\propto\sum_i \gamma_i(\Delta_i)$Re$\left(E/\sqrt{E^2-\Delta_i^2}\right)$, convoluted with the derivative of the Fermi function at the corresponding temperature. For a perfectly isotropic s-wave BCS superconductor with gap $\Delta_{BCS}$, $\gamma_i(\Delta_i)=1$ when $\Delta_i=\Delta_{BCS}$ and zero elsewhere. In systems with an anisotropic band structure, however, the superconducting gap is often not perfectly homogeneous and there is a certain distribution of gap values over the Fermi surface\,\cite{PhysRevB.92.054507,Herrera2023,PhysRevResearch.4.023241,PhysRevLett.101.166407,PhysRevB.97.134501,GarciaTalavera2025}.
\subsection{Estimation of critical field and coherence lengths}
As we discuss in the Supplemental Material\,\cite{SM_1}, we measure the upper critical field using the magnetic field dependence of the tunneling conductance in between vortices and the magnetic field dependence of the radial behavior of the tunneling conductance around vortices, finding  $H_{c2}\approx$1.8\,T and an in-plane coherence length of about $\xi_{a,b}=13$~nm.
 
There is no signature of superconductivity in bulk measurements in the field and temperature range discussed in Figs.\,\ref{fig:gapvst},\,\ref{fig:vortices}. The anisotropy of the upper critical field ($\xi_{a,b}\sim 20 \, \xi_{c}$)\,\cite{Veyrat2023} shows that superconductivity in $\gamma-$PtBi$_2$ is confined to a thin surface layer. The very short coherence length along the c-axis suppresses proximity effects from the normal bulk, leaving the surface density of states essentially unaffected.
\\
\\
Note: During the final stages of the preparation of this work, we have become aware of an independent similar work which reports the observation of bulk superconductivity with $T_C \sim$0.35~K and surface superconductivity with $T_C \sim $3~K. Only atomically flat surfaces are found, without an observable vortex lattice under magnetic fields \cite{Zhang25}.

\section*{Data Availability}
The data generated in this study have been deposited in the osf.io database Ref. \cite{osf}

\clearpage
\newpage

\section*{Supplementary Material}

\section{Superconducting properties in different fields of view}

\subsection{Characterization of the corrugated surface areas on flakes}

\begin{figure*}[ht!]
	\centering
		\includegraphics[width=0.9\linewidth]{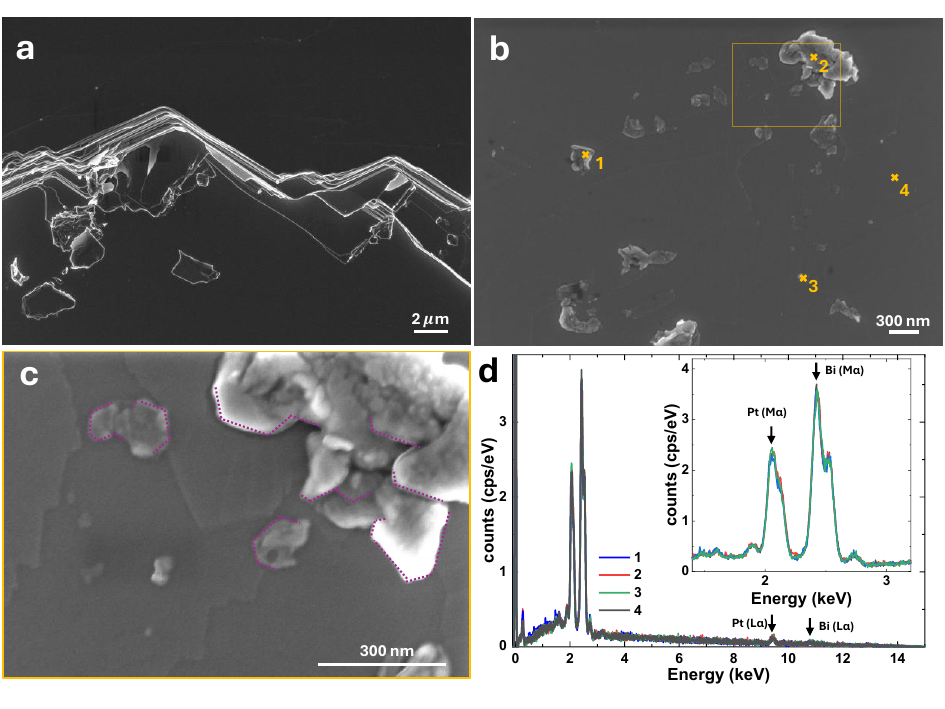} 
	\caption{(a) Scanning Electron Microscope (SEM) image of the surface of $\gamma$-PtBi$_2$ at an edge of the sample. White scale bar is 2~$\mu$m long. Notice the presence of small flakes which have been caused during cleavage. (b) SEM image at the interior of the sample. We can see a large flat surface and several flakes lying on the flat. We have performed a compositional analysis using Energy Dispersive electron Spectroscopy (EDS). We find a homogeneous composition at sites marked as 1-4 in (b) which coincides with the stoichiometry of $\gamma-$PtBi$_2$ within 0.05 uncertainty in the Pt-Bi ratio. Sites 1-3 are on different flakes, and 4 is on a flat terrace. Results in other parts of the sample are similar. In (c) we show a zoom on the ocre rectangle shown in (b). Dashed purple lines are drawn with 120$^{\circ}$ vertices to highlight the hexagonal shape of flakes. Notice that, whereas the terrace is fully flat, the flakes has a strong corrugation, suggesting that there are often compositions of several flakes. White scale bar is 300~nm long in (b,c). (d) EDS spectra measured at the points marked in (b). Inset shows a zoom around the M lines of Pt and Bi. We mark the position of M and L lines of Pt and Bi with arrows.} 
		\label{fig:SEM}
\end{figure*}

As we show in the main text, we find areas which are atomically flat, and also areas with a certain corrugation. The corrugated areas are located on top of the atomically flat areas. We name these in the following nanometer-sized flakes for reasons that will become clear below.

We have imaged the surface of our cleaved samples after STM measurements using high-resolution Scanning Electron Microscopy (SEM) to characterize the origin and composition of the nanometer-sized flakes. We have used an Electron Beam Lithography eLINE-PLUS from Raith GmbH, which incorporates a Bruker XFlash 6I30 EDS detector. In Fig.\,\ref{fig:SEM} we show several SEM images of different sizes. We can see on the edge of the sample (Fig.\,\ref{fig:SEM}(a)) that the step edges often have 120$^{\circ}$ corners, reflecting the trigonal symmetry of the crystal lattice. Furthermore we find flakes close to the edges. These flakes have probably been caused during cleaving and we can find them all over the surface, including flat areas well inside the sample (Fig.\,\ref{fig:SEM} (b)). The flakes have sizes similar to those that we have observed in our STM measurements. In Fig.\,\ref{fig:SEM}(c) we show a high resolution SEM image on the area marked by an ocre rectangle in Fig.\,\ref{fig:SEM}(b). We see that all flakes have borders with an hexagonal shape, marked in Fig.\,\ref{fig:SEM}(c) by purple dashed lines. However, their surface is not completely flat, contrasting the surface of the rest of the sample. There are flakes with a large surface corrugation (upper right part of Fig.\,\ref{fig:SEM}(c)) and others with much less corrugation (left part of Fig.\,\ref{fig:SEM}(c)). This agrees remarkably well with our STM observations---flakes that are one or a few unit cells high and are relatively flat (as those discussed below in Fig.\,\ref{fig:hilera}(a,c,d) and in Fig.\,\ref{fig:Agujero}(d,g)), and flakes showing surfaces with enhanced corrugation. 

To analyze the elemental composition, we have performed Energy Dispersive electron Spectroscopy (EDS) measurements on some of these flakes (sites 1-3 marked in in Fig.\,\ref{fig:SEM}(b)) and on top of the terrace (site 4 in Fig.\,\ref{fig:SEM}(b)). Results are shown in Fig.\,\ref{fig:SEM} (d). There is no difference between spectra performed on the flakes (sites 1-3) and on flat terraces (site 4). A standardless quantitative analysis yields ratios of Pt/Bi atomic concentration of 0.46 (site 1), 0.49 (site 2), 0.46 (site 3) and 0.46 (site 4). Within the experimental uncertainty of 0.05 in the Pt/Bi ratio, this coincides with the stoichiometric ratio of $\gamma-$PtBi$_2$. We note that the electron–sample interaction volume may introduce a small uncertainty in these measurements. From the measurement parameters (working distance = 10.3~mm, accelerating voltage = 15~kV), we estimate an electron beam radius of approximately 400 nm and an interaction depth of about 700~nm. The lateral beam size is sufficiently small that the beam falls entirely within most flakes. However, the interaction depth implies that, for sites 1–3, the EDS signal is expected to include a contribution from the underlying flat terrace beneath the flakes. Nevertheless, if these flakes were a species different from PtBi$_2$, we would expect the corresponding EDS spectra to differ at least qualitatively from that measured at site 4.

\begin{figure*}[ht!]
	\centering
		\includegraphics[width=0.8\linewidth]{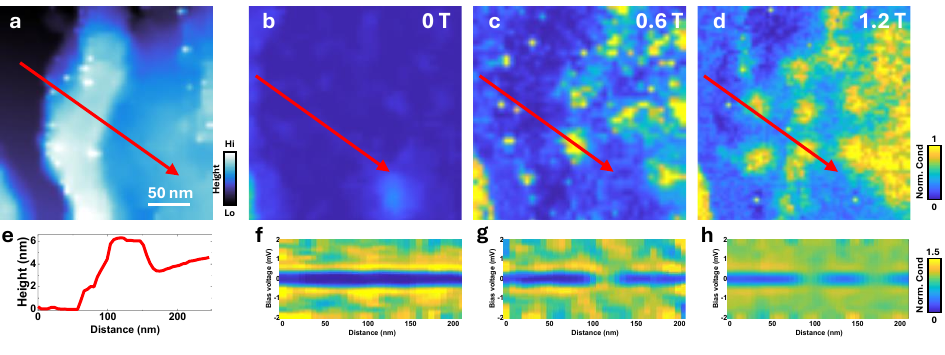} 
	\caption{(a) STM topography of a field of view comprising an atomically flat surface and a nanometer-sized flake. White scale bar corresponds to 50~nm. Color scale is shown to the right. Zero-bias tunneling conductance map of the vortex lattice in the same field of view as (a) as a function of magnetic field at 0~T (b), 0.6~T (c) and 1.2~T (d). (e) Height profile along the path marked with a red arrow in (a). (f-h) Conductance profile along the path marked as a red arrow in (b-d). The gap size remains when crossing from the atomically flat surface onto the nanometer-sized flake and no boundaries can be observed on the profiles. Data taken at 100~mK.}
		\label{fig:fieldprofiles}
\end{figure*}

\begin{figure*}
	\centering
		\includegraphics[width=1\linewidth]{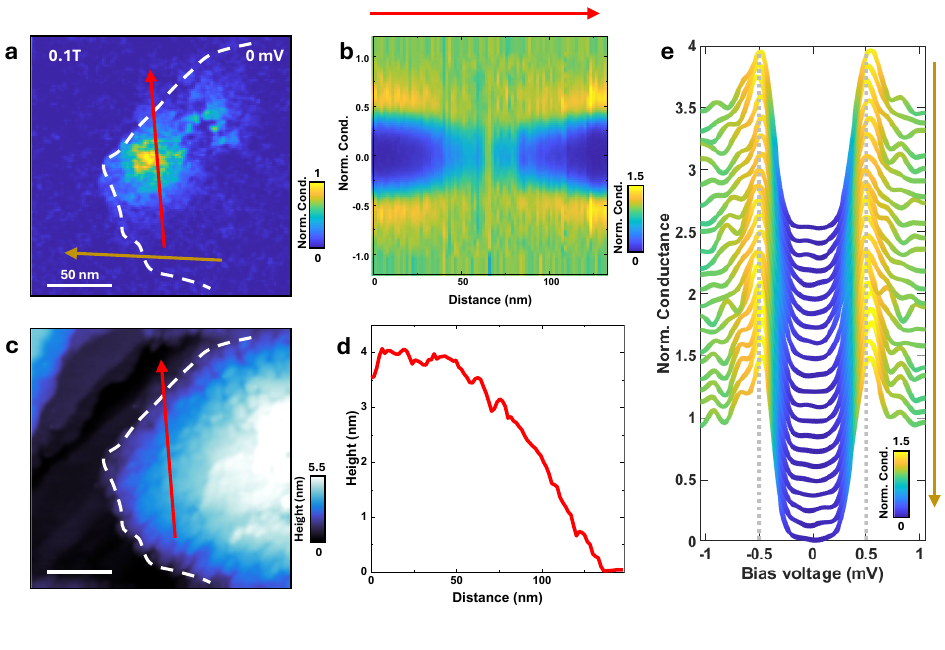} 
	\caption{(a)  Zero-bias conductance map obtained on a single vortex. Dashed line shows the contour separating a nanometer-sized flake (right side of the dashed line) from flat terraces (left side of the dashed line). Color scale is shown on the bottom right. White scale bar is 50\,nm long. (b) Tunneling conductance vs bias voltage along the red arrow in (a). (c) STM topography of the same field of view as in (a). Note the presence of atomic steps on the top left corner, having no significant influence on the zero bias tunneling conductance. (d) Height profile along the red arrow in (c). We set the height at zero on the atomically flat area. For reference, the c-axis unit cell is 0.617~nm large. (e) Conductance curves along the orange line in (a). Curves have been shifted vertically for clarity. We mark the position of quasiparticle peaks with grey dotted lines.  Data taken at 100~mK.}
	\label{fig:rough}
\end{figure*}

\subsection{Superconducting properties in nanometer-sized flakes, compared to those in flat areas}

We show topographies and conductance maps in fields of view that contain atomically flat surfaces and nanometer-sized flakes in Fig.\,\ref{fig:fieldprofiles}, as a function of magnetic field, and in Fig.\,\ref{fig:rough} in a different field of view. Topographies are shown in Fig.\,\ref{fig:fieldprofiles}(a) and Fig.\,\ref{fig:rough}(c), and the corresponding zero-bias conductance map are shown in Figs.\, \ref{fig:fieldprofiles}(b-d) and Fig.\,\ref{fig:rough}(a). In Figs.\,\ref{fig:fieldprofiles}(f-h), and Figs. \,\ref{fig:rough}(b,e) we show the conductance profiles crossing the boundary between flat surfaces and nanometer-sized flakes. Nanometer-sized flakes present a superconducting density of states which is equivalent to the one observed in atomically flat surfaces. We observe the same gap size, and coherence peaks with the same shape at zero field and under magnetic fields (see also Fig.\,3(h,g) in the main text).

\section{Vortex lattice observations with STM}
\subsection{Mechanisms for absence of vortices on flat surfaces}
In the following, we describe a possible scenario in which the apparent absence of vortices on atomically flat surfaces may be explained by enhanced vortex mobility. Such enhanced mobility could arise in a very weak pinning regime, either through thermal activation or via electrostatic interactions between the STM tip and the vortices.

\subsubsection{Vortex pinning}

Assuming vortices are present on atomically flat terraces, the pinning force $F_{p}$ can be estimated by considering the free energy per unit volume from the thermodynamic critical field and the vortex volume\,\cite{PhysRevLett.58.599}. Due to the tubular structure of vortices, the pinning force depends critically on the vortex length. Estimates of the pinning force per unit vortex length are often of order of $10^{-6}$~N/m\,\cite{PhysRevLett.58.599,Zhang15}. In $\gamma-$PtBi$_2$, the bulk is normal, which implies that superconductivity is restricted to a small layer close to the surface and vortices are essentially two-dimensional pancakes. Taking a conservative estimate for the thickness for the superconducting layer of 1 nm (approximately two unit cells along the c-axis, unit cell size along the c-axis is of 0.617~nm), we estimate $F_{p1}\sim 1 $~fN \cite{PhysRevB.43.7837}. For comparison, the pinning strength is similar to the weak pinning observed in trilayer graphene and below the one observed in disordered thin films\,\cite{mahapatra2025quantumcriticalitytunablegriffiths,PhysRevB.103.214512} (the pinning strength in units of the thermal energy is given further below). 

To delve into this aspect, we can estimate the possible shift of vortex positions $\delta$ by considering collective pinning in two dimensions\,\cite{Larkin1979}. There is a region where the vortex lattice is correlated, of size $R_c$. The elastic energy density is then given by $\frac{1}{2}C_{66}\left(\frac{\delta}{R_c}\right)^2$ in three dimensions, where $C_{66}$ is the elastic shear modulus. In two dimensions, the energy is independent on $R_c$, and is given by $\Delta E \approx \frac{1}{2}C_{66}\left(\frac{\delta}{R_c}\right)^2 \pi R_c^2 d\approx C_{66}\delta^2 d$, where $d$ is the thickness of the superconducting layer. We can estimate $C_{66}$ using $C_{66}\approx \frac{H_c^2}{16 \pi}b(1-b)^2$\cite{Brandt1977}, with $H_c$ the thermodynamic critical field and $b=\frac{H}{H_{c2}}$. For magnetic fields at about half the upper critical field and for external forces in the fN range, we have $F\approx \frac{\Delta E}{\delta}$, and we find a vortex shift which is of a few times the coherence length $\xi_{a,b}$. Therefore, the two-dimensional vortex lattice is very soft.

On flakes, the vortex pinning should increase. We have observed flakes above 5 nm in height, roughly an order of magnitude larger than the c-axis unit cell. The eventual increase in vortex length should lead to an increased pinning force $F_{p2}$. For example, in flakes that are close to a c-axis unit cell size, vortices seem unpinned while they remain static on flakes with larger thickness, as we show below in Fig. \ref{fig:hilera}. 

\subsubsection{Tip-vortex interaction and thermal effects}

\begin{figure*}[ht!]
	\centering
		\includegraphics[width=0.9\linewidth]{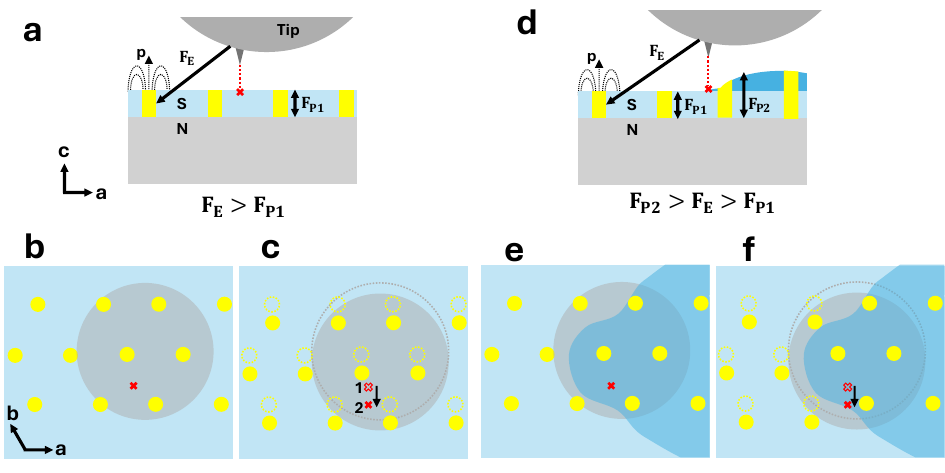} 
	\caption{(a) Schematic lateral view of the STM measurement of the vortex lattice on a flat defect-free region. Superconducting topmost layer is shown as light blue, the bulk normal metal as light grey and the tip as dark grey. Vortices are in yellow. The red cross represents the position where the tunneling is measured. Black arrows represent schematically the electrostatic interaction between the tip and a vortex ({\bf F$_E$}) and the pinning force ({\bf F$_P$}). We schematically represent the electric dipole {\bf $p$} on vortices\cite{Blatter96} as a dashed arrow. (b,c) Vortex lattice in two instances of the STM sweep on an atomically flat surface. The tip moves from top to the bottom following the black arrow in (c). Due to the interaction between the tip and the vortex lattice, vortices are dragged by the tip. (d) Schematic view of the same arrangement as in (b,c), but now with an area on a nanometer-sized flake at the right. There are now two pinning forces, the one on the nanometer-sized flake ({\bf F$_{P2}$}) being larger than the one on flat surfaces ({\bf F$_{P1}$}). (e,f) Two instances of the tip's motion. Vortices on the nanometer-sized flake remain fixed and do not move with the tip, while those lying on atomically flat areas move.}
		\label{fig:esquema}
\end{figure*}

 We illustrate schematically the possible interaction between tip and vortex lattice in Fig.\,\ref{fig:esquema}. As is well known, a STM tip is most often blunt, but there is always a small atomic size apex which dominates in the tunneling process, because of the exponential dependence of the tunneling current with distance\,\cite{PhysRevB.31.805}. We represent the tunneling apex by a small triangular-like tip in Fig.\,\ref{fig:esquema}. The tunneling current from this apex is shown schematically in red. The overall shape of the tip is blunt and we represent it by a semi-spherical object with a radius of several nm. The electrostatic interaction between tip and sample occurs over a much larger scale than tunneling. This length scale depends on the size of the tip but is at least several nm in diameter\,\cite{PhysRevLett.95.136802,PhysRevB.70.073312,Ast2016,PhysRevB.88.035436,PhysRevLett.70.2471,Guillamon2021}. The electrostatic interaction between tip and dipoles on the sample contains several components, such as image charges and the actual dipolar interaction\,\cite{PhysRevLett.95.136802,PhysRevB.88.035436,PhysRevLett.70.2471}. The latter can be, in principle, attractive and repulsive, depending for instance on the sign of the bias. Independently on the sign of the interaction, for a certain electrostatic configuration, there is an equilibrium distribution of vortices below the tip, most often consisting of the vortex lattice below the capacitor built by the spherical tip (with nm or tens of nm radius). The charge of these vortices can eventually influence the electrostatic tip-sample interaction, particularly for two-dimensional vortices.

As shown in Ref.\,\cite{Blatter96}, the differences in the chemical potential between superconducting and normal quasiparticles induce a charge redistribution in vortices. Metallic screening does not fully eliminate the charge. This results in an electric dipole $p$ oriented perpendicular to the surface whose strength is of the order of $p\sim e d$ where $e$ is the elementary charge and $d\sim 1$\,\AA (which gives $p\sim1$ D, one Debye). Such a minute electric dipole has been often considered previously and has an influence in several properties of superconductors, such as the Hall effect in the mixed state\,\cite{PhysRevLett.75.1384,doi:10.1021/acs.nanolett.1c04688,MACHIDA2002443,Faure_2007,PhysRevB.105.064514}. To estimate the force generated by the electric field of the tip $E$ on the vortex, we first consider the corresponding interaction energy,  $U\sim E \,p$. The dipole sits on the vortex core, but its field extends over a distance of order of the superconducting coherence length $\sim \xi_{a,b}$. We consider a tip which is at a distance smaller than $\xi_{a,b}$ and is blunt and large in size. The effective electric field is then $E \sim V/\xi_{a,b}$, where $V$ is the voltage applied to the tip. We can then estimate the force acting on the vortex as $F_E\sim U/\xi_{a,b}\sim Vp/\xi_{a,b}^2$. For $V\sim 0.1$\,V, $\xi_{a,b}\sim10$\,nm we obtain $F_E\sim10$\,fN. These estimations suggest that the electrostatic force could be greater than the pinning force on atomically flat surfaces $F_{p1}$.

In practice, the interaction could just lead to changes in the position of a few vortices, because the hexagonal vortex lattice is not perfect and the vortex positions are slightly influenced by small defects. With very weak pinning, assuming $F_{p1}<F_E$, we obtain the result shown schematically in Fig.\,\ref{fig:esquema}(b,c). When we move the tip the equilibrium position of some vortices move together with the tip in such a way as to maintain the distribution of charges below the tip. In Fig.\,\ref{fig:esquema}(d) we show an increase in the effective length of the magnetic vortex in nanometer-sized flakes. This could increase the pinning force, eventually leading to $F_{p2}>F_E$. In that case, we have the situation shown in  Fig.\,\ref{fig:esquema}(e,f). The vortex lattice on the flakes remains fixed (dark blue on Fig.\,\ref{fig:esquema}(e,f), $F_{p2}>F_E$), but vortices on the flat two-dimensional surface are moved by the interaction with the tip (light blue on Fig.\,\ref{fig:esquema}(e,f), $F_{p1}<F_E$).

These estimations suggest that on large and flat surfaces, $F_{p1}<F_E$, vortices could be highly mobile and remain undetected. However, on flakes of sufficient thickness, $F_{p2}>F_E$, the vortex lattice could be pinned and STM might then reveal the presence of vortices.

Alternatively, thermal effects can also play a relevant role in a weak pinning regime. The pinning strength, estimated by taking $U_p\approx F_{p1} d$, with $d$ the thickness of the superconducting layer, is of $U_p\approx 0.1$~K for $d\approx 1$~nm. Thermal excitations, even at the low temperatures we are measuring, are in the range of the estimated pinning interaction, and could thus lead to relevant vortex mobility. Vortex thermal fluctuations on flakes are discussed further below.

In any case, both effects discussed above could play a cooperative effect on enhancing the mobility of vortices. An enhanced mobility of vortices on atomically flat surfaces is compatible with observations in Fig.\,3(g,h) of the main text where we find that conductances curves on atomically flat surfaces are equivalent to those taken on flakes. We provide experimental results suggestive of thickness-dependent pinning of vortices and tip-induced vortex motion on nanometer-sized flakes further below.

The possible electrostatic interaction between tip and sample complements previous STM work in semiconducting materials, where such interactions essentially lead to band-bending and changes in the lateral resolution\,\cite{PhysRevB.88.035436,PhysRevLett.70.2471}. In all cases, electrostatics leads to an invasive interaction. In superconductors, the electric field is most often perfectly screened and such interactions have been neglected. But on vortices such effects could have the a-priori unexpected consequence of allowing for enhanced vortex mobility in sufficiently thin superconductors. In any event, we remark that an electrostatic interaction does not strongly depend on the tip-sample distance as far as the latter is measured using STM. A decrease in the tunneling current (exponential dependence with distance) by an order of magnitude leads to a change of the tip-sample distance by just one \AA, which leads to small changes in an electrostatic force (quadratic dependence with distance). Absence of vortices has been repeatedly found in STM experiments under magnetic field in $\gamma$-PtBi$_2$ using values of the tunneling current which vary from 100 pA to several nA \cite{Schimmel2024,Hoffmann2024,Zhang25}.

Absence of vortices has been also reported previously in highly disordered and very thin films particularly close to the superconductor to insulator transition\,\cite{PhysRevB.88.014503,PhysRevResearch.2.033307,PhysRevB.105.L140503}. An enhanced vortex mobility and tip-sample interactions could play a certain role in those observations too, as well as in other two-dimensional superconducting systems\,\cite{Cao2018,Han2025,mahapatra2025quantumcriticalitytunablegriffiths,Kezilebieke2020,Deng2024,PhysRevB.103.214512,Lechermann2024,fan2025,Ming2023}.

The charge of vortices carrying Majorana modes has been investigated theoretically. Depending on the particular superconducting state, vortices carrying Majorana fermions are expected to present a considerably enhanced charge with respect to conventional s-wave superconductors\,\cite{Masaki_2018,PhysRevB.99.054512}. Our data suggest that the vortices on atomically flat terraces carry a substantial charge, which might favor the interaction with the tip.

\subsection{2D behaviour of vortices}
\subsubsection{Enhanced vortex mobility}

\begin{center}
\textbf{Thickness-dependent pinning}
\end{center}

\begin{figure*}[ht!]
	\centering
		\includegraphics[width=0.9\linewidth]{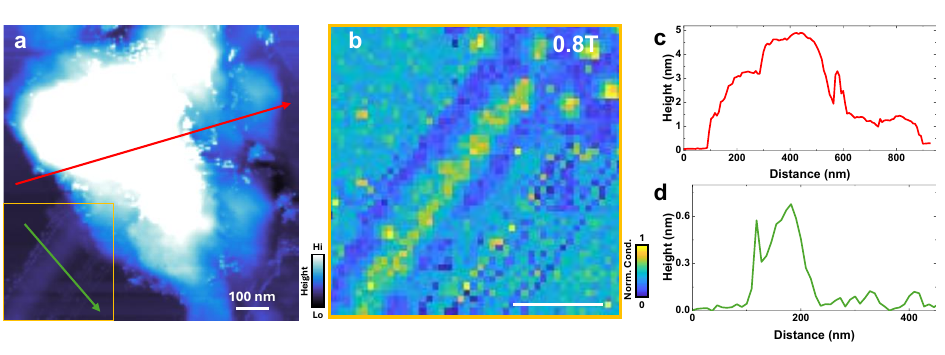} 
	\caption{(a) STM topography of the same field of view shown in Fig. 3 (c,f) of the main text. White line corresponds to 100\,nm in size. (b) Zero-bias tunneling conductance map within the orange square shown in (a) at 0.8\,T. This is the left bottom part of the figure shown in Fig.\,3(e) of the main text. We see a row of a finite density of states, which we attribute to vortices moving along an elongated nanometer-sized flake. (c,d) Profiles of the topography corresponding along the red (c) and green (d) arrows shown in (a). We note that the height of the nanometer-sized flake along (d) is an order of magnitude smaller than the one shown in (c), enabling vortex mobility on the corresponding area.  Data taken at 100~mK.}
		\label{fig:hilera}
\end{figure*}

Of interest is also a closer analysis of the field of view shown in Fig.\,3(c,d) of the main text. In Fig.\,\ref{fig:hilera}(a) we show the topography acquired simultaneously with the tunneling conductance map shown in  Fig.\,3(c) of the main text. In the center of the field of view there is a large nanometer-sized flake. On the bottom left corner (orange square) there is a stripe-like flake separating two atomically flat regions. In Fig.\,\ref{fig:hilera}(b) we show the tunneling conductance map corresponding to this area. We see that individual vortices are barely observed, and instead there is a central line with a large density of states. It is important to note that this striple-like nanometer-sized flake is smaller in height than the one present on the center of Fig.\,\ref{fig:hilera}(a). The height profiles, shown in Fig.\,\ref{fig:hilera}(c,d) show a size difference of approximately an order of magnitude. The height of the one-dimensional part of the flake is roughly one unit cell along the c-axis (0.619~nm). This implies that the vortex length is smaller on the stripe-like nanometer-sized flake, which allows for vortex mobility and produces a one-dimensional stripe of mobile vortices.

Thus, there seems to be a gradual evolution with increasing flake height from the highly mobile vortices on flat surfaces to the completely pinned vortices on large nanometer-sized flakes. This suggests that vortices on flakes are longer than those on the atomically flat areas, enhancing their pinning, similar to pancake vortices in cuprate superconductors\,\cite{doi:10.1126/science.aat1773,doi:10.1143/JPSJ.72.2153}. We also note that vortices seem to generally align with the edges of the flakes. The vortex Fig.\,\ref{fig:rough}(a) is located at an edge of the nanometer-sized flake. Furthermore, the vortex lattices shown in Fig.\,3(b,c,d) of the main text and in Fig.\,\ref{fig:vorticeswithtemperature} are oriented following the edges of the flakes.

\begin{center}
\textbf{Tip-vortex interaction on nanometer-sized flakes.}
\end{center}

\begin{figure*}[ht!]
	\centering
		\includegraphics[width=0.95\linewidth]{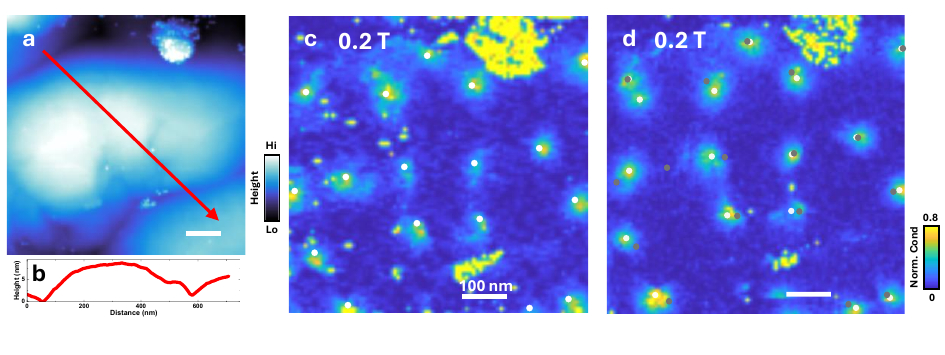} 
	\caption{ (a) STM topography on a large nanometer-sized flake. (b) Height profile along the red arrow shown in (a). (c,d) Consecutive zero-bias conductance maps in the same field of view at 0.2~T. White scale bar corresponds to 100~nm. Color scale is shown to the right. We mark vortex positions with white dots. In (d) we  mark the position of vortices in (c) with grey dots, evidencing that vortices have moved, possibly, as a consequence of the tip-sample interaction. Tunneling current was above 2~nA, approximately three times the current used in other maps.  Data taken at 100~mK.}
		\label{fig:drag}
\end{figure*}

\begin{figure*}[ht!]
	\centering
		\includegraphics[width=0.95\linewidth]{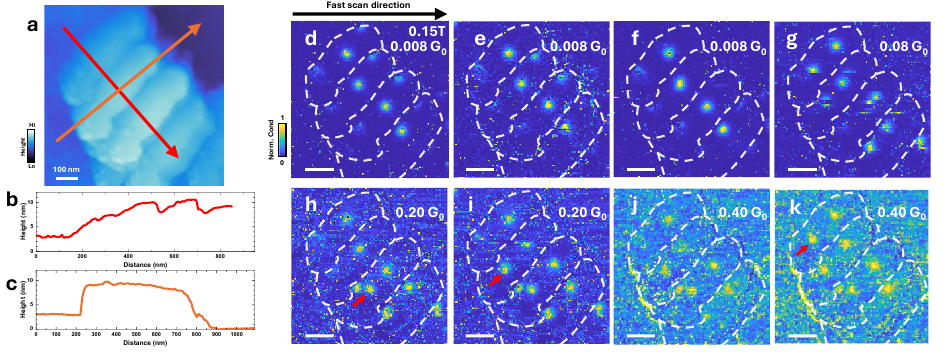} 
	\caption{ (a) STM topography on a large nanometer-sized flake. (b)((c)) Height profile along the red (orange) arrow shown in (a). (d-g)  Consecutive zero-bias conductance maps in the same field of view at 0.15~T (intervortex distance is of $\sim$150~nm at this magnetic field) performed with increasing tunneling current. Tunneling conductance is given in units of the quantum of conductance $G_0=2e^2/h\approx77.5 \mu$S: Voltage is fixed to 1.6~mV and tunneling current is 1~nA (d-f), 10~nA (g), 25~nA (h,i), 50~nA (j,k). White scale bar corresponds to 100~nm. Color scale is shown to the left. We mark the boundary of different flakes with dashed white lines. In (h-k) we further mark with red arrows the position of vortices where we observe motion after successive measurements in the same field of view.  Data taken at 100~mK.}
		\label{fig:dragloop}
\end{figure*}

In Fig.\,\ref{fig:drag} we show results on a field of view containing a large single flake with an applied magnetic field of 0.2~T, which is relatively low (far from $H_{c2}$). This image has been taken with a tunneling current of about 2~nA. Both features, increased tunneling current (decreased tip-sample distance) and low field, provide a stronger tip-sample interaction. In Fig.\,\ref{fig:drag}(a,b) we show the topography and a line scan. In Fig.\,\ref{fig:drag}(c,d) we show two zero bias conductance images taken in the same field of view, (d) taken immediately after (c). White dots in Fig.\,\ref{fig:drag}(c,d) provide the vortex positions in each scan. In Fig.\,\ref{fig:drag}(d) we show as grey points vortex positions of Fig.\,\ref{fig:drag}(c). We see that vortices are slightly displaced.

Further evidence for vortex displacements is presented in Fig. \,\ref{fig:dragloop}. In this field of view we observe several nanometer-sized flakes (magnetic field of 0.15~T). In Fig.\,\ref{fig:dragloop} (a-c) we show the topography and two line scans across flakes and atomically flat surfaces. In Fig.\,\ref{fig:dragloop}(d-f) we show zero bias conductance images taken consecutively in the same field of view. The tunneling conductance at high bias voltages is of 0.008$G_0$. Vortices remain stable, except perhaps at the edges of the flakes. When we increase the tunneling conductance by one order of magnitude, (0.08$G_0$ Fig.\,\ref{fig:dragloop}(g)) we observe vortex instabilities created by the scanning tip, as shown by the horizontal cuts observed in some vortices in the zero bias tunneling conductance. When further increasing the tunneling conductance (Fig.\,\ref{fig:dragloop}(h-k), at 0.2$G_0$ and 0.4$G_0$) we observe additional vortices in the flakes (for example those marked by red arrows in Fig.\,\ref{fig:dragloop}(h,i,k)). Repeated scanning (Fig.\,\ref{fig:dragloop}(h,i) and Fig.\,\ref{fig:dragloop}(j,k)) leads to small changes in the vortex configuration. In the latter images, we often observe two vortex rows, as opposed to one vortex row at the start of the experiment (Fig.\,\ref{fig:dragloop}(d)). This apparent increase in the vortex density could be due to vortices dragged from other areas into the field of view, or to thermally excited vortices which fall into strong pinning centers with time.

The evolution of the vortex configuration shown in Figs. \ref{fig:drag},\ref{fig:dragloop} confirms an enhanced vortex mobility. This, together with the thickness-dependent pinning in Fig. \ref{fig:hilera}, suggests that (assuming the tip-vortex interaction described above) the pinning force acting on nanometer-sized flakes $F_{p2}$ could be of the same order as the electrostatic interaction between STM tip and vortices, $F_E$. By reducing the distance between tip and sample at working at a low applied magnetic field, our results suggest an enhanced tip-vortex interaction. Alternatively, vortices might be simply thermally excited and their position change slowly with time among different sites with some help from a small interaction with the tip.

\subsubsection{Temperature dependence of the vortex lattice}

\begin{figure*}[ht!]
	\centering
		\includegraphics[width=1\linewidth]{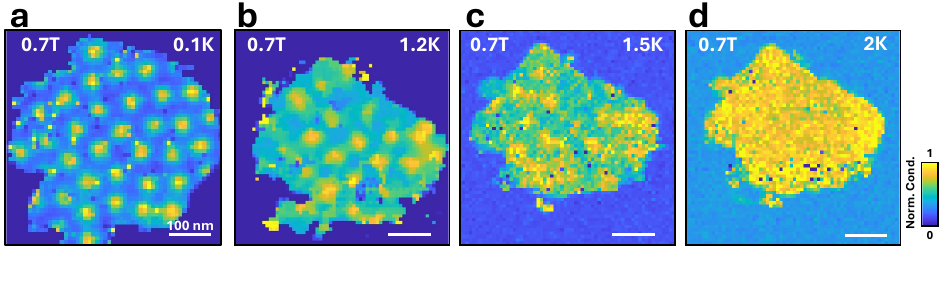} 
	\caption{Zero-bias tunneling conductance map of the vortex lattice at 0.7~T in the same field of view as a function of temperature at 0.1~K (a), 1.2~K (b), 1.5~K (c), and 2~K (d). The nanometer sized flake where we can observe the vortex lattice decreases its size due to interactions with the tip (see also Fig. \ref{fig:Agujero}). White scale bars are shown at the bottom and correspond to $100$~nm in all images. The color scale is shown at the right and is the same for all images. We observe that the vortices increase considerably their mobility with increasing temperature.}
		\label{fig:vorticeswithtemperature}
\end{figure*}

We show the effect of increasing temperature on the vortex lattice of $\gamma-$PtBi$_2$ in Fig.\,\ref{fig:vorticeswithtemperature}(a-d). The scans are made consecutively and in the same field of view. The nanometer-sized flake reduces its size with subsequent scans, possibly due to tip-sample interactions (similar tip-sample interactions are shown and discussed in Fig. \ref{fig:Agujero}). The surface of the flake remains large enough to identify the vortex lattice at the applied magnetic field. Vortices are observed only within the nanometer-sized flake. The hexagonal vortex lattice remains distinguishable up to 1.2~K, but is more disordered as compared to the vortex lattice at lower temperatures. At 1.5~K, there are practically no traces of superconducting vortices. However, the average zero bias tunneling conductance normalized to its value at high voltages within the flake is well below one. This suggests that the flake is still superconducting. However, vortices strongly fluctuate in their position, giving a time averaged tunneling conductance which is neither one, as observed in a pinned vortex core, nor very low, as observed in between vortices\,\cite{Guillamon2009,PhysRevB.88.014503,PhysRevB.105.L140503}. At 2~K, the zero bias tunneling conductance on the nanometer-sized flake is homogeneous, suggesting strongly thermally activated vortex motion and a melted vortex lattice\,\cite{Guillamon2009,Duhan2025,PhysRevB.108.L180503,PhysRevB.105.L140503,PhysRevB.103.214512}. Remarkably, the tunneling conductance in the flat areas outside the nanometer-sized flake strongly increases with temperature too. This suggests that the thermal jitter of vortices induces a finite tunneling conductance also on the flat areas.

\subsubsection{Vortex behaviour with magnetic field history}
The vortex lattice on flakes is observed with similar properties when reversing the direction of the magnetic field. Furthermore, zero field cooled and field cooled experiments lead to similar observations too. This is to be expected due to the strong field penetration in two-dimensional superconductors.

\subsection{Tip interaction with nanometer-sized flakes}   

\begin{figure*}[ht!]
	\centering
		\includegraphics[width=1\linewidth]{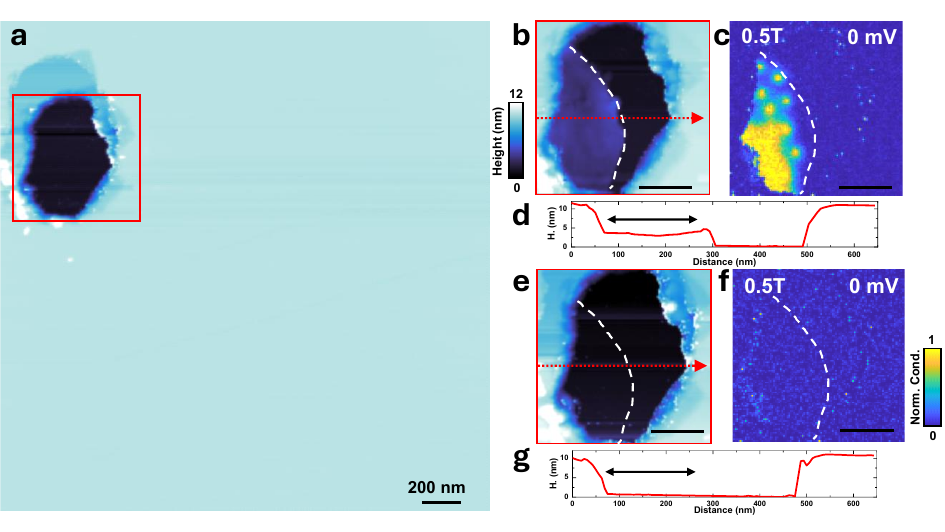}
		\vskip 9mm
		\caption{(a) Large scale STM topography of an atomically flat area (2.5\,$\mu$m x 2.5\,$\mu$m). Notice the presence of a hole like structure on the top left part of this area. The hole is approximately 10 nm deep. (b) STM topography image of the top left area of (a). Within the hole, we find a nanometer-sized flake. The white dashed line shows the separation between flake (left side) and the atomically flat terrace (right side) within the hole. (c) Tunneling conductance map at zero bias in the same field of view as in (b). We observe a few vortices on the nanometer-sized flake. The intervortex distance is as expected for the applied magnetic field (0.5~T). (d) Height profile along the red dashed line in (b). (e) STM topography map of the same area as in (c), but after several scans. We see now that the surface within the hole is atomically flat. The interaction between tip and sample removed the rough area shown in (b). (f) Tunneling conductance map at zero bias on the same field of view as in (e). We can no longer identify vortices, showing that the vortex lattice in the whole field of view is now mobile. (g) Height profile along the red dashed line in (e).  Data taken at 100~mK.}
		\label{fig:Agujero}
\end{figure*}

Another example of a large atomically flat surface at an applied magnetic field of 0.5~T is shown in Fig.\,\ref{fig:Agujero}(a). This surface presents a ``hole'' on the upper left corner, with an approximate depth of 10~nm. In Fig.\,\ref{fig:Agujero}(b) we see the result of a first scan. Inside the hole we find a nanometer-sized flake. Although the superconducting gap is perfectly homogeneous over the whole surface, we see the vortices on the nanometer-sized flake within the hole Fig.\,\ref{fig:Agujero}(c). After making several scans, we find the topography shown in Fig.\,\ref{fig:Agujero}(e). The surface inside the hole is now perfectly flat, because interactions with the tip have modified the position of the nanometer-sized flake, possibly removing it altogether from this field of view. We then see no trace of the vortices in Fig.\,\ref{fig:Agujero}(f) and instead a spatially homogeneous superconducting tunneling conductance with the properties discussed in the main text.

Sometimes we observe a tunneling conductance which does not vary smoothly with position. For example, the tunneling conductance varies rapidly between light blue and dark blue along the two sides of the white dashed line of Fig.\,4(a) of the main text, or at the vortices close to the edges in Fig.\,4(b) of the main text. This could suggest that vortices are ``truncated'' at the edges of flakes. We believe that this is due to a tip effect. When having a sharp transition between surfaces of different roughness, the tip could actually change slightly. These effects often occur in presence of strong changes of corrugation, as crossing from an atomically flat surface to a corrugated surface. They do not imply distortions in the measured topography by more than a few nm. However, when the tunneling conductance strongly varies within a few nm, distortions are enough to produce changes in the tunneling conductance.

\section{Upper critical field and vortex core size as a function of the magnetic field}

\begin{figure*}[ht!]
	\centering
		\includegraphics[width=0.7\linewidth]{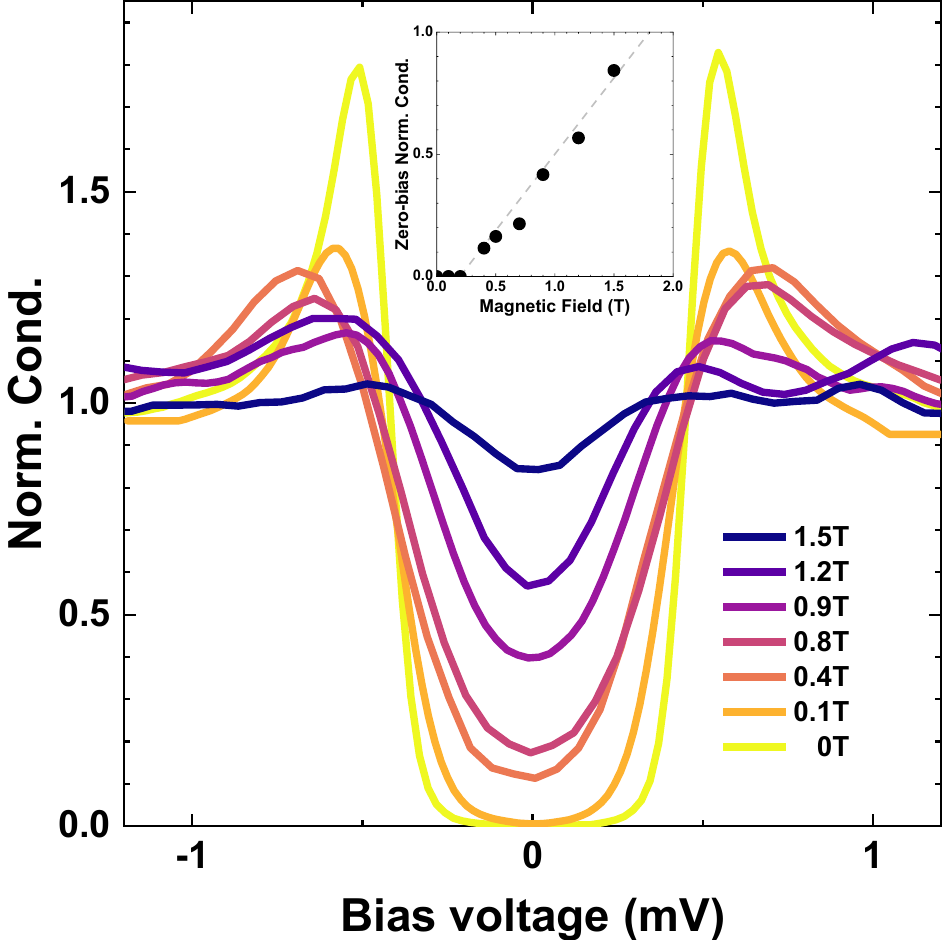} 
	\caption{Tunneling conductance curves as a function of applied magnetic field. Curves are obtained in between vortices on flakes at a temperature of 100 mK. Inset shows the zero-bias normalized tunneling conductance for several fields. Dashed grey line is a guide to the eye and follows linearly the increase in conductance. It reaches a normalized conductance of 1 at $H\sim 1.8$~T. }
		\label{fig:gapvsB}
\end{figure*}

In Fig.\,\ref{fig:gapvsB} we show the normalized tunneling conductance curves as a function of the magnetic field, taken in between vortices on top of flakes. We find suppressed coherence peaks and a gradual filling of the gap when increasing the magnetic field. To estimate the value of the upper critical field we trace the zero bias conductance as a function of the magnetic field (inset of Fig.\,\ref{fig:gapvsB}) and find that it reaches the normal state value at a magnetic field of approximately $H_{c2}\approx 1.8$\,T.

\begin{figure*}[ht!]
	\centering
		\includegraphics[width=0.7\linewidth]{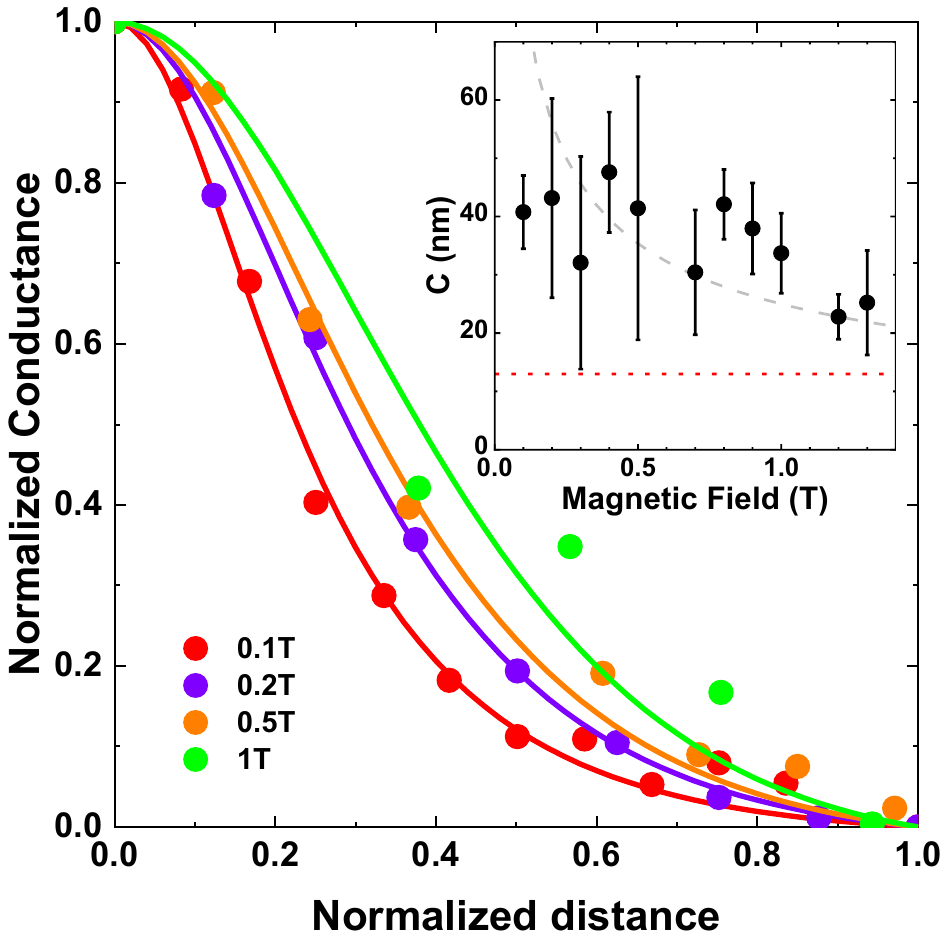} 
	\caption{In the main panel we show by colored points the radially-averaged tunneling conductance normalized to its value in and outside vortex cores, following Ref.\,\cite{PhysRevB.94.014517}, for the values of the magnetic field shown in the legend. The colored lines are fits to the model of Ref.\,\cite{PhysRevB.94.014517}. In the inset we show as dots the magnetic field dependence of the vortex core size $C$, defined as in Ref.\,\cite{PhysRevB.94.014517}. The horizontal dotted line shows $\xi_{a,b}\sim$13\,nm. Dashed grey line indicates the expected vortex core size behaviour with field $\propto 1/\sqrt{H}$ for a superconductor in the clean limit\,\cite{PhysRevB.94.014517}.}
		\label{fig:coresize}
\end{figure*}

The magnetic profile of vortices expands in a two-dimensional superconductor\,\cite{10.1063/1.1754056}. But the spatial profile of the superconducting order parameter around the vortex core does not depend on the dimensionality and remains of the order of the coherence length perpendicular to the magnetic field, $\xi_{a,b}$. There is a characteristic evolution of  the profile of the superconducting density of states around a vortex with magnetic field. When the vortex density increases with magnetic field, the slope of the superconducting order parameter $\Delta$, given by $d\,\Delta/d\,r |_{r \to 0}$ eventually presents a magnetic field dependence which was considered in Ref.\,\cite{PhysRevB.71.134505,PhysRevB.94.014517}. $d\,\Delta/d\,r |_{r \to 0}\propto 1/C(H)$, with $C$ being the so-called vortex core size which coincides exactly at the upper critical field $H_{c2}$ with $\xi_{a,b}$. It was shown that the radial dependence of the tunneling conductance $\sigma$ can be written as:

\begin{equation}
\sigma=1-\frac{\rho^2 (1+\eta^2)}{\rho^2+\eta^2}\exp\left( \frac{\eta^2 (1-\rho^2)}{1+\eta^2} \right)
\label{modelokogan}
\end{equation}

where $\rho=1.05 r/L$, $\eta=1.05 C/L$, and $L$ is half the intervortex distance. The model relates the density of states with the tunneling conductance using deGennes model\,\cite{deGennes1964} and assumes a smooth overlap on the order parameter value between different vortices at $d\,\Delta/d\,r|_{r = L/2}=0 $. In Fig.\,\ref{fig:coresize} we show the vortex core size $C$, obtained from fitting the radially-averaged tunneling conductance along vortex cores to equ.\,\ref{modelokogan}. We find that the vortex core size $C$ is slightly decreasing with the magnetic field and tends towards the coherence length $\xi_{a,b}=13$~nm at the critical field $H_{c2}=1.8$~T. This behavior is expected for a superconductor with an electronic mean free path of order of $\xi_{a,b}$\,\cite{PhysRevB.94.014517}.

\section{Josephson coupling between $\gamma$-PtBi$_2$ and superconducting tips}

\begin{figure*}[ht!]
	\centering
		\includegraphics[width=0.9\linewidth]{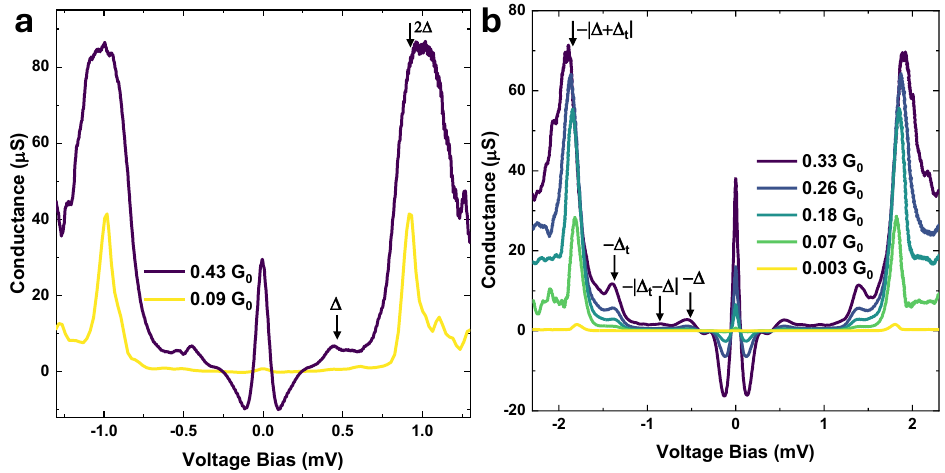} 
	\caption{(a) Tunneling conductance obtained after picking a lose piece of $\gamma$-PtBi$_2$ with the tip. (b) Tunneling conductance obtained with a tip of Pb. The normal state tunneling conductance is provided in the legend in units of the quantum of conductance $G_0=\frac{2e^2}{h}=7.748 10^{-5}$~S. Black arrows mark features related to the gap of the different electrodes and are discussed in the text.  Data taken at 100~mK.}
		\label{fig:josephson}
\end{figure*}

To obtain further evidence of the robustness of the superconducting properties of $\gamma$-PtBi$_2$ we performed controlled indentations of the tip into the sample, picking up pieces of $\gamma-$PtBi$_2$. We observed the Josephson effect and traced its evolution when changing the tunneling current. We observed similar behavior as found previously in systems such as Pb or Al\,\cite{Rodrigo2004}. There are peaks in the tunneling conductance at twice the superconducting gap and at the superconducting gap, due to Andreev reflection. In addition, there is a pronounced peak at zero bias which evidences the passage of a Cooper pair (Josephson) supercurrent (Fig.\,\ref{fig:josephson}(a)). When measuring with a tip of Pb, we also observe a  Josephson peak (Fig.\,\ref{fig:josephson}(b)) and the features in the tunneling conductance characteristic of tunneling between superconductors with different gap sizes\,\cite{PhysRevB.74.132501}, marked by arrows in Fig.\,\ref{fig:josephson}(b).

\section{Further aspects of the STM experiment}

We cleaved samples in-situ at 4.2~K or below. Samples were cleaved by gluing a post of alumina on top of the samples and hitting the alumina by moving the sample holder below a beam. We measured hundreds of fields of view in four different samples, obtaining the behavior discussed in the main text in each of them. 

We used tips of Pt-Ir, except in the experiment with superconducting tip, where we used Pb tips. To prepare and clean the tips we used the tip preparation methods described in Ref.\,\cite{Rodrigo2004,JGRodrigo2004}.

We would like to remark that we usually find a single surface termination when measuring a certain sample of $\gamma$-PtBi$_2$. To obtain data from the two surface terminations, we first cleaved one sample and measured it, finding atomically flat terraces with termination A (Fig.\,1 (b) of the main text). The cleaved piece was attached to a rope and was let to fall down to the bottom of the inner vacuum chamber of the dilution refrigerator. We then recovered this part and mounted it in such a way as to make sure that the upper surface was at the other side as compared with previous experiment. We then cooled and cleaved the sample again. We found, in that new cleave, always atomically flat terraces presenting topographies as shown in Fig.\,1(c) of the main text, as opposed to the topographies found in the original cleave (Fig.\,1(b) of the main text). Therefore, these two types of images present the two types of surfaces obtained from a cleave, A and B in Fig.\,1(b,c) of the main text. The superconducting behavior was the same for both surface terminations.

Flakes do not present large atomically flat surfaces, but instead inclined planes and height changes that are of atomic size or below, but do not form a periodic spatial pattern. This suggests that flakes are not completely flat and consist of many terraces (see also Scanning Electron Microscopy (SEM) results above). We note, however, that we can find flakes of one unit cell height, as shown in Fig.\,\ref{fig:hilera}(a,d), and others with about eight unit cells, as shown in Fig.\,\ref{fig:Agujero}(d).

During scanning, the current was most often below 1~nA and the bias voltage of a few mV. We increased the tunneling current until 50~nA, observing the same features (absence of vortex lattice in atomically flat areas and presence in corrugated areas). We usually present the tunneling conductance normalized to its value well above the superconducting gap edge. Tunneling conductance maps were obtained by cutting the feedback loop at each pixel on the scan area and making a full tunneling conductance vs bias voltage measurement.

Sometimes it is useful to fit the tunneling conductance to a lifetime broadened density of states\,\cite{PhysRevLett.41.1509,PhysRevB.94.144508}. This often allows discussing broadened quasiparticle peaks. In our case, this does not lead to appropriate fits, because the tunneling conductance deviates from zero very rapidly when increasing the finite lifetime in the model, whereas it remains very small in the experiment. Therefore, we cannot fit at the same time broadened quasiparticle peaks and the low energy tunneling conductance.

\section{Single crystal growth of $\gamma-$PtBi$_2$ and bulk characterization}

Elemental Bi and Pt were combined
in a Bi80-Pt20 ratio and placed in a fritted, alumina crucible set (available as Canfield Crucible
Set or CCS)\,\cite{Canfield02012016,CCS}. The CCS was sealed into an amorphous silica tube and placed into
a box furnace\,\cite{Canfield_2020}. The ampoule was then heated to 1000~$^{\circ}$C over 5 hours, held at 1000
$^{\circ}$C for 5 hours, cooled to 700 $^{\circ}$C over 4 hours and then slowly cooled to 440 $^{\circ}$C over 146
hours. At 440~$^{\circ}$C, above the reported 420~$^{\circ}$C transformation temperature to the cP12/Pa-3
phase of PtBi$_2$ [ASM phase diagram entry 979935 for Bi-Pt binary phase diagram]\,\cite{Okamoto1991}, the ampoule
was removed from the furnace and the excess, Bi-rich solution was decanted\,\cite{Canfield_2020}. Once the
ampoule cooled to room temperature it was opened and large, cleanly faceted plates of hP9/P-3 (space group no.\,157) PtBi$_2$ were revealed. X-ray diffraction measurements taken
ground single crystals confirmed the
correct, high temperature phase and further illustrated the stability of the phase at room temperature\,\cite{oleary2025}.

\begin{figure}[ht!]
	\centering
		\includegraphics[width=0.95\linewidth]{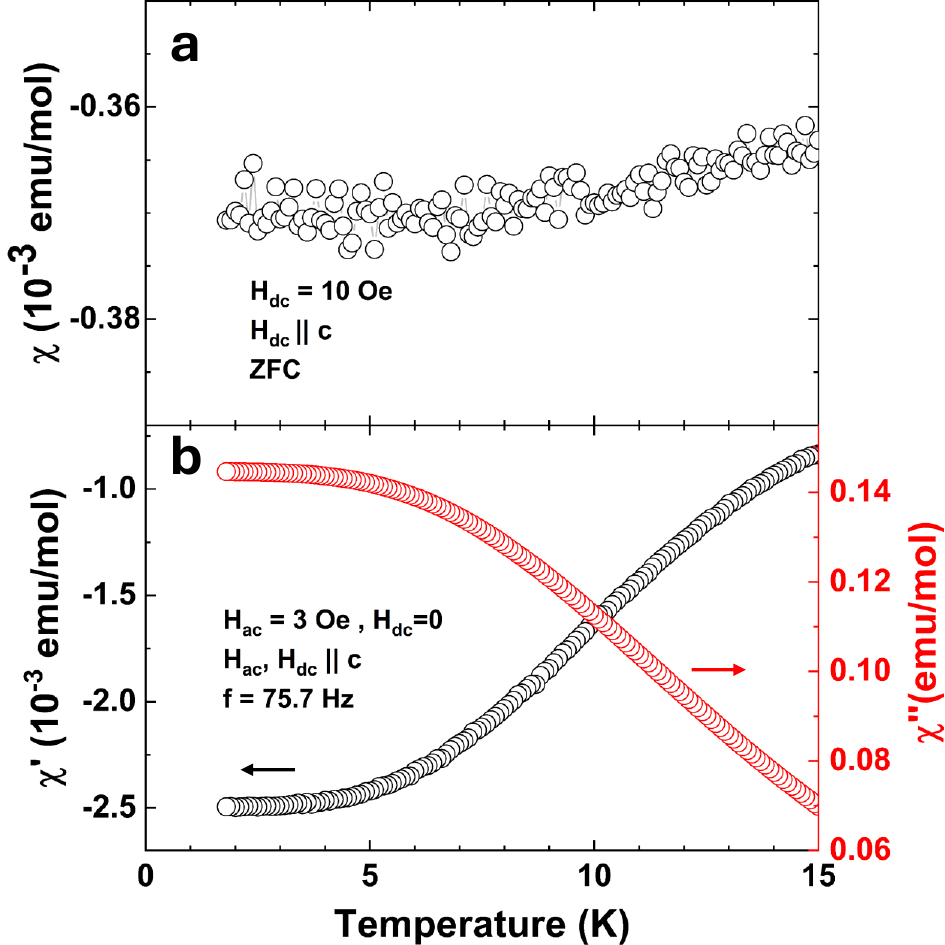} 
	\caption{(a) Temperature dependence of the magnetic susceptibility $\chi$ of a $\gamma-$PtBi$_2$ single crystal. The applied field is 10~Oe along the c-axis. (b) Temperature dependence of the real ($\chi'$) and imaginary ($\chi''$) part of the ac magnetic suspectibility. The applied field is 3 Oe along the c-axis and the frequency is of 75.7~Hz.}
		\label{fig:susceptibility}
\end{figure}

We show measurements of the zero field cooled magnetization and the AC susceptibility as a function of temperature in our single crystals in Fig.\,\ref{fig:susceptibility}. To maximize any possible trace of superconductivity in the magnetization, we used samples with a plate-like form, oriented perpendicular to the magnetic field (c-axis parallel to the magnetic field). As we can see in Fig.\,\ref{fig:susceptibility}, there are no traces of superconductivity.

\section{DFT calculations}

Band structure of $\gamma-$PtBi$_2$ in space group 157 with spin-orbit coupling (SOC) in DFT\,\cite{PhysRev.136.B864,PhysRev.140.A1133} have been calculated with PBE3\,\cite{PhysRevLett.78.1396} exchange-correlation functional, a plane-wave basis set and projected augmented wave method\,\cite{PhysRevB.50.17953} as implemented in VASP\,\cite{PhysRevB.54.11169,KRESSE199615}. A Monkhorst-Pack\,\cite{PhysRevB.13.5188} ($8\times 8 \times 8$) k-point mesh with a Gaussian smearing of 0.05 eV including the $\Gamma$ point and a kinetic energy cutoff of 230.3 eV have been used. A tight-binding model based on maximally localized Wannier functions\,\cite{PhysRevB.56.12847,PhysRevB.65.035109,RevModPhys.84.1419} is constructed to reproduce closely the bulk band structure including SOC in the range of E$_F\pm2$ eV with Pt $sd$ and Bi $p$ orbitals. The surface spectral function and the QPI pattern are shown as the joint density of states\,\cite{doi:10.1126/science.aad8766} of the semi-infinite surface. These have been calculated with the surface Green’s function methods\,\cite{MPLopezSancho1984,MPLopezSancho1985} as implemented in WannierTools\,\cite{WU2018405}.

The cleaving energies calculated within PBE+D3\,\cite{PhysRevLett.78.1396,10.1063:1.3382344} for each of the planes of the crystal structure along the c-axis, shown in Fig.\,1(a) (middle panel) are of 2.66~J/m$^2$ for Pt-Bi$_2$, 3.01 J/m$^2$ for Pt-Bi$_3$ and 0.89 for Bi$_1$-Bi$_3$. This agrees with the structural parameters, which provide an interlayer Pt-Bi$_3$ distances of 1.33\,\AA, but an interlayer Bi$_1$-Bi$_3$ distance of 2.36~\AA. Therefore, the cleaving plane is in between the Bi$_1$ and Bi$_3$ layers, giving the two surfaces shown in Fig.\,1(b,c) of the main text.

\section{Comparison with the superconducting properties found in literature}

Previous bulk resistivity measurements have found a transition to the superconducting state at $\sim$0.6~K-1.1~K\cite{shipunov2020,Zabala2024}. However, the critical current and the critical field were very small. No trace of superconductivity is found in the specific heat down to 0.5 K~\cite{Schimmel2024}. Thin flakes with a thickness of the order of tens of nanometers have shown a Berezinskii-Kosterlitz-Thouless transition at $\sim300$~mK\,\cite{Veyrat2023}.  Superconductivity with $T_c$ up to $\sim2.5$~K appears with doping with Rh, Se and Te\cite{shipunov2020,doi:10.7566JPSJ.91.034703} or by applying hydrostatic pressure\,\cite{wang2021}.

On the other hand, techniques probing the surface have found a plethora of different critical temperatures ranging from $\sim3$~K (point-contact)\,\cite{bashlakov2022}, to $\sim6$~K (SQUID on tip)\,\cite{Schimmel2024} and to $\sim$8-14~K\,\cite{Kuibarov2024} (ARPES). Most remarkably, STM measurements found superconducting gaps compatible with a $T_c$ of the order of $\sim$80~K\cite{Schimmel2024,besproswanny2025temperaturedependencesurfacesuperconductivity} and a $H_{c2}>15$~T. However, superconducting properties probed in Ref.\,\cite{Schimmel2024} were inhomogeneous\,\cite{Hoffmann2024}. Furthermore, the vortex lattice could not be observed in Ref.\,\cite{Schimmel2024} despite having performed measurements under high magnetic fields.

Other work in $\gamma$-PtBi$_2$ did not find evidence for superconductivity in STM\,\cite{guo2025} and ARPES (down to 3~K)\,\cite{oleary2025}. This is compatible with the very recent experiment of Ref.\cite{Zhang25}, where bulk superconductivity with a very small upper critical field and vortices whose size is an order of magnitude larger than the ones we report here, were observed below 0.35~K. Furthermore, a surface superconducting density of states below 2.9~K was observed in atomically flat areas similar to those we report in our experiment, again without observing the two-dimensional surface vortex lattice.
\bibliography{biblio}

\end{document}